\documentclass{article}
\usepackage{spconf,amsmath,graphicx}
\usepackage{booktabs}
\usepackage[textfont=it,tableposition=top]{caption}
\usepackage{hyperref}

\title{Parallel WaveNet conditioned on VAE latent vectors}
\name{J Rohnke, T Merritt, J Lorenzo-Trueba, A Gabrys, V Aggarwal, A Moinet, R Barra-Chicote}
\address{Amazon.com}
\begin{document}
\ninept
\maketitle
\begin{abstract}
Recently the state-of-the-art text-to-speech synthesis systems have shifted to a two-model approach: a sequence-to-sequence model to predict a representation of speech (typically mel-spectrograms), followed by a 'neural vocoder' model which produces the time-domain speech waveform from this intermediate speech representation. This approach is capable of synthesizing speech that is confusable with natural speech recordings. However, the inference speed of neural vocoder approaches represents a major obstacle for deploying this technology for commercial applications. Parallel WaveNet is one approach which has been developed to address this issue, trading off some synthesis quality for significantly faster inference speed. In this paper we investigate the use of a sentence-level conditioning vector to improve the signal quality of a Parallel WaveNet neural vocoder. We condition the neural vocoder with the latent vector from a pre-trained VAE component of a Tacotron 2-style sequence-to-sequence model. With this, we are able to significantly improve the quality of vocoded speech.
\end{abstract}
\vspace{-0.1cm}
\section{Introduction}
\vspace{-0.1cm}
Statistical parametric speech synthesis (SPSS) systems were developed to enable a consistent level of naturalness which is flexible across the wide range of linguistic contexts which may be encountered at synthesis-time \cite{zen2009statistical,zen2015acoustic}. In this paradigm statistical models predict parameters of speech from linguistic context features. A time-domain speech waveform is constructed from these parameters using a signal processing-based vocoder \cite{kawahara2001aperiodicity,morise2016world}. Whilst SPSS systems are flexible and produce a consistent level of naturalness, the vocoders used at the time place a ceiling effect on the naturalness that is possible within this synthesis paradigm \cite{merritt2017overcoming}. Hybrid synthesis aimed to leverage the stability of statistical models in order to drive the selection of units in a unit selection system \cite{qian2012unified,merritt2016deep,wan2017google}, thus removing the effect of signal processing-based vocoders. However, hybrid synthesis still relies on the selection units observed in the training data. This makes joins inevitable; due to the large number of possible linguistic contexts it is impossible for these to all be present in the speech database.

Neural approaches have been proposed to overcome the aforementioned ceiling effect of vocoding. Originally WaveNet-like approaches \cite{oord2016wavenet,arik2017deep} predicted the time-domain waveform signal from linguistic context and f0, using a large model of stacked dilated convolutions and autoregression. This model is effectively being asked to estimate the vocal tract configuration in addition to constructing the time domain waveform. More recently, researchers in literature have found that breaking the problem down into 2 stages allows the models to produce very high quality audio \cite{shen2018natural}: 1) estimate a speech representation (e.g., mel-spectrograms), 2) use a neural vocoder to produce the time domain waveform from the speech representation. In addition, this makes the task of the speech waveform prediction component simpler, enabling for more compact models to produce high quality speech \cite{Kalchbrenner2018,Lorenzo-Trueba2018}. Although it has been possible to make neural vocoding models more compact whilst preserving high quality, one issue which remains is the time required for inference. This is largely due to the presence of autoregression in the model topologies and is a limitation when putting neural vocoders into production for commercial systems.
 
Approaches aiming to solve this limitation can be grouped into three main categories: knowledge distillation-based (e.g. Parallel WaveNet \cite{VanDenOord2018} and ClariNet \cite{Ping2019}), directly training the data likelihood maximization (e.g. WaveGlow \cite{Prenger2019}), and adversarial training (e.g. WaveGAN \cite{Donahue2019}). In this investigation we focus on the Parallel WaveNet neural vocoding approach due to its success in commercial deployments.

Parallel WaveNet was proposed to improve the inference speed at synthesis-time. It uses a large WaveNet neural vocoder model as a teacher. A student model is then trained to replicate the predictions of the teacher network. The student network learns to shape noise input into a speech waveform based on the conditioning passed to the vocoder. Crucially, the student network uses Inverse Autoregressive Flows (IAFs) \cite{Kingma2016} for which the sampling procedure is easy to parallelize because all noise samples are available at the same time and thus vastly improves the inference speed. 

Even though Parallel WaveNet improves the performance speed compared to neural vocoders, whilst maintaining a high signal quality, these approaches are still behind the quality of natural recordings. We train our vocoders on natural recordings to solve the spectrogram to waveform problem. During inference we use it to synthesize from predicted spectrograms, introducing a mismatch due to the imperfect predicted spectrograms. To help bridge this gap, we train the network with additional conditioning in the form of a pre-trained Variational Auto Encoder (VAE) \cite{Kingma2014} latent vector. This utterance level vector is extracted from the VAE reference encoder of a Tacotron 2-like sequence-to-sequence acoustic model \cite{Zhang2019LearningLR}. We show on two different speakers that this additional conditioning improves synthesis quality for predicted spectrograms, especially for very expressive voices. During inference, the conditioning vector can be replaced by a constant vector from within the training distribution without loss of quality, resulting in efficient inference without the need for additional calculations.

\vspace{-0.3cm}
\section{Architecture} \label{sec: architecture}
\vspace{-0.1cm}
As described in the introduction, we are working in the paradigm of having two separate models for synthesis: an acoustic model that predicts mel-spectrograms from text and a neural vocoder that predicts audio from the predicted mel-spectrograms.

\vspace{-0.1cm}
\subsection{Neural Vocoder} \label{sec: pawa}
\vspace{-0.2cm}
For the neural vocoder, we use a model inspired by Parallel WaveNet. This means we train two models using the Student-Teacher paradigm. The teacher is a WaveNet-like model which provides sequential sampling but parallel likelihood computation via teacher forcing which makes it fast to train but slow for inference. The student uses IAFs where the affine transform parameters are predicted by autoregressive conditioners built from WaveNet-like blocks of dilated convolutions. IAFs allow parallel sampling which makes them efficient during inference. But they require sequential calculations to estimate likelihood and are therefore slow to train. Training first the teacher and then the student to match the teacher's outputs,  results in a model that can be trained and sampled from efficiently.

\begin{figure}[t]
  \centering
  \includegraphics[width=\linewidth]{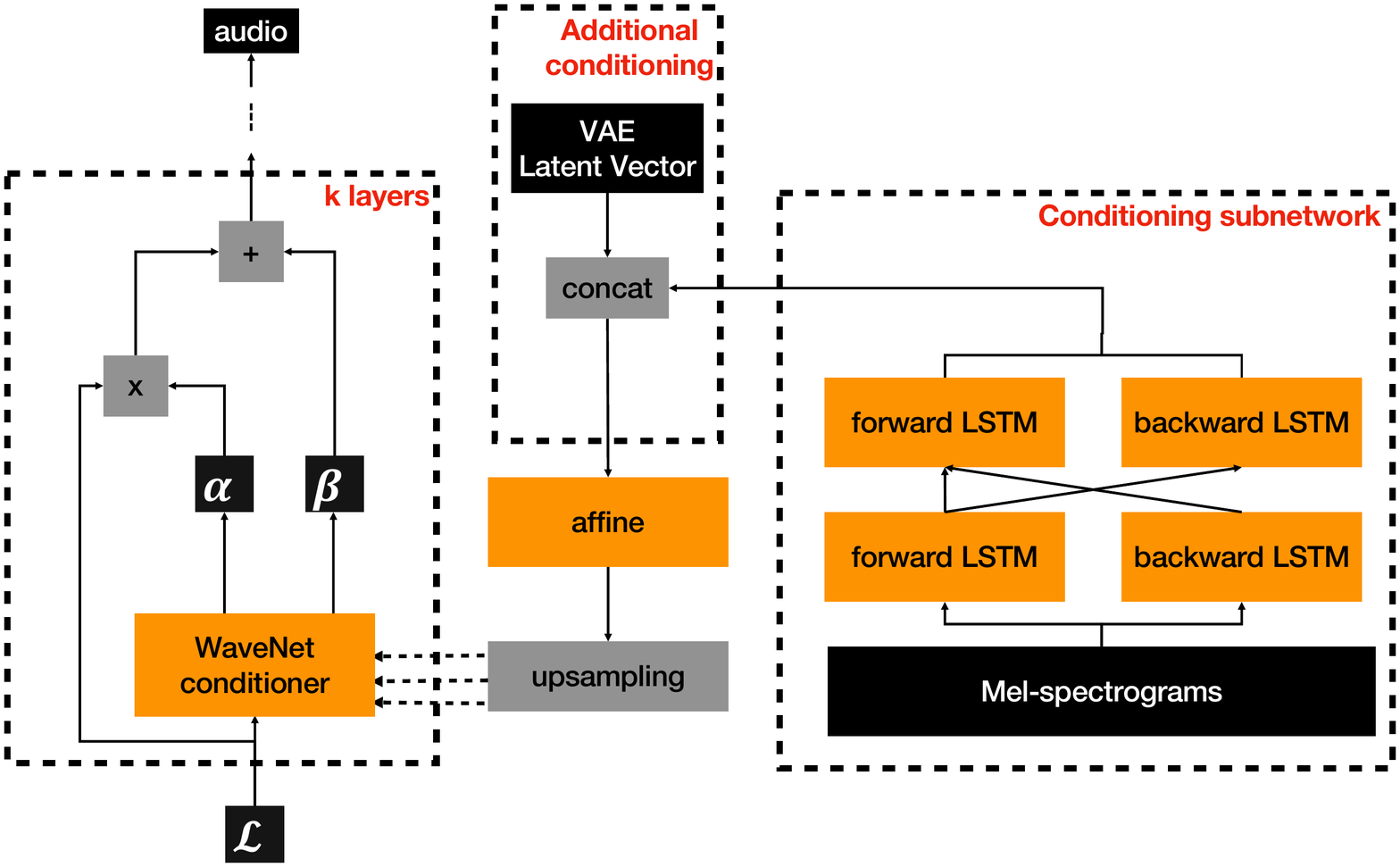}
  \vspace*{-1.1cm}
  \caption{The architecture of the Parallel WaveNet-like neural vocoder. The additional conditioning block adds the conditioning on the latent representation extracted from a VAE reference encoder of a Tacotron 2-like acoustic model.}
  \vspace{-0.7cm}
  \label{fig:pawa_diagram}
\end{figure}

The WaveNet-like models used for both student and teacher are convolutional networks which are conditioned on textual information via extra bias parameters in the convolutional layers. The original paper proposed the use of linguistic features, f0 and durations. Instead, we condition on mel-spectrograms. As observed in the original paper, Student-Teacher training alone leads to whispering artefacts. Hence, the authors proposed training on a combination of other losses: Power Loss, Phoneme Recognition Style Loss, and Contrastive Loss. We additionally propose conditioning both teacher and student models on the utterance-level latent vector extracted as described in section \ref{sec: prosotron}. The latent vector is concatenated with the output of the conditioning sub-network (see Figure~\ref{fig:pawa_diagram}). 

\vspace{-0.3cm}
\subsection{Latent Vector Extractor} \label{sec: prosotron}
\begin{figure}[t]
  \centering
  \includegraphics[width=\linewidth]{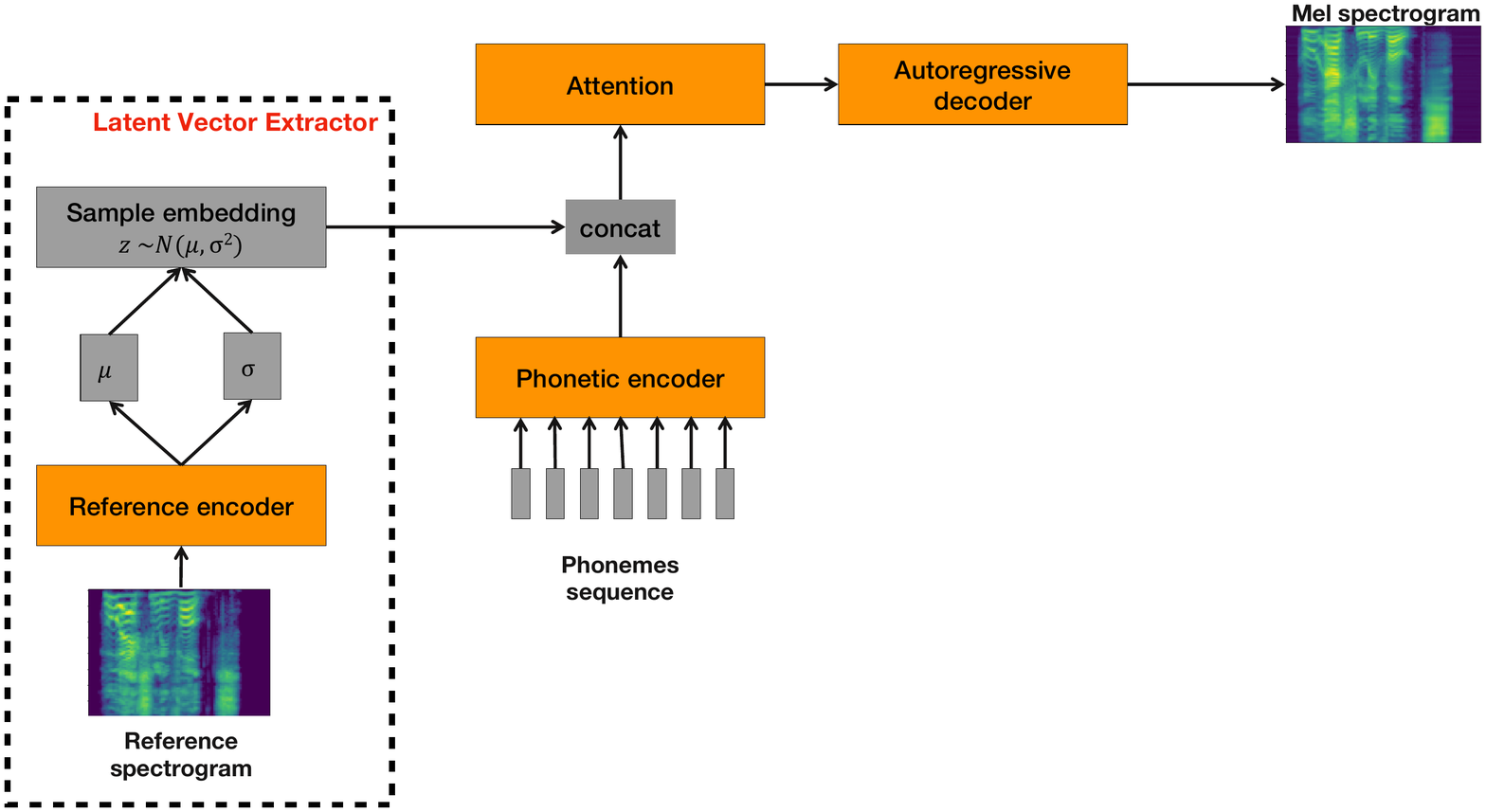}
\vspace{-1cm}
  \caption{The architecture of the acoustic model and the latent vector extractor used to condition the neural vocoder.}
\vspace{-0.6cm}
  \label{fig:acoustic_model_diagram}
\end{figure}

\vspace{-0.2cm}
To extract the utterance-level latent vector we use to condition the Parallel WaveNet models, we train an acoustic model with a Tacotron 2-like architecture \cite{Zhang2019LearningLR} (see Figure~\ref{fig:acoustic_model_diagram}). It uses a VAE reference encoder \cite{hodari2019using} to extract a latent representation from the target mel-spectrograms. This VAE reference encoder consists of a stack of convolutional layers followed by a bi-LSTM and two projections for the mean and standard deviation of an n-dimensional Gaussian distribution from which an n-dimensional latent vector is sampled.\\
Mel-spectrograms (extracted from recordings or predicted by an acoustic model) are input to the VAE reference encoder block to extract an utterance-level bottlenecked representation. We are effectively repurposing the VAE reference encoder trained alongside the acoustic model as a latent vector extractor that is used to condition the neural vocoder.

\vspace{-0.3cm}
\section{Experiments} \label{sec: experiments}
\vspace{-0.1cm}

\subsection{Data and model details} \label{sec: models}
\vspace{-0.2cm}
We train the acoustic model, the Parallel WaveNet teacher and student models on a diverse set of $63$~hours of data from a single female US speaker. It contains $46$~hours of audio in the traditional neutral style and additional $17$~hours of expressive data covering the following domains: news \cite{Prateek2019}, long-form book reading\footnote{https://developer.amazon.com/en-US/blogs/alexa/alexa-skills-kit/2020/04/new-alexa-long-form-speaking-style-and-polly-voices}, as well as excited and disappointed emotions\footnote{https://developer.amazon.com/en-US/blogs/alexa/alexa-skills-kit/2019/11/new-alexa-emotions-and-speaking-styles}.

The acoustic model architecture is similar to \cite{Prateek2019} with the exception of replacing the two-dimensional style ID with a VAE reference encoder that outputs a $64$-dimensional latent vector. This places less constraints on the categories of speech samples since the representation is extracted from the signal rather than a manual label which is potentially inconsistent. The VAE latent space is learned in an unsupervised manner without providing explicit information about the different expressive domains. We follow \cite{Zhang2019LearningLR} and increase the Kullback-Leibler divergence (KLD) loss factor from $0$ to $1$ between iterations $25$k and $150$k and additionally only take it into account every $200$ training steps to avoid the well-documented KL loss collapse. Once this model is converged, we use the VAE reference encoder to extract latent vectors for all the training data. These are used as additional conditioning for the training of the Parallel WaveNet teacher and student models.

The Parallel WaveNet teacher model uses a hidden size of $128$ for the two-layer bi-LSTM in the conditioning subnetwork. The output of this is concatenated with the $64$-dimensional VAE latent vector extracted from the acoustic model, before being upsampled to match the audio. The main block of the teacher model is a WaveNet architecture with $30$ layers of dilated convolutions where the dilation value is doubling every layer and reset to $2$ every $10$ layers. The Residual Gated CNN blocks use $256$-dimensional skip and gated channels, the filter activation is tanh and the gate activation is sigmoidal. The output is a mixture of $10$ logistics. The Parallel WaveNet student model shares the conditioning subnetwork with frozen weights with the teacher. The student is using $4$ affine transform flow layers conditioned using a WaveNet like architecture predicting scale and shift. The flow conditioners contain dilated convolutions with $10$, $10$, $10$ and $30$ layers. The last block uses same dilation value growth and reset as the teacher network. The dilated convolutions use $64$-dimensional gated channels with tanh filter activation and sigmoidal gate activation. We are training the vocoder using a batch size of $16$, the Adam optimizer with default parameters and learning rate decay of $0.95$ for the teacher and no decay for the student. We use Polyak averaging with decay of $0.999$. 

\vspace{-0.2cm}
\subsection{In-domain evaluation}
\vspace{-0.2cm}
To evaluate the quality of our proposed model, we conduct a preference test comparing it to a baseline. We use mel-spectrograms predicted by the acoustic model described in section \ref{sec: prosotron} which also functions as the latent vector extractor. The baseline model is a Parallel WaveNet student as described in section \ref{sec: pawa}, i.e. without additional conditioning, trained on the data described in section \ref{sec: models}. Our proposed model differs from the baseline only in having the additional conditioning on the latent vector during training and inference. We evaluate $464$ utterances covering all expressive domains described in section \ref{sec: models} by asking $30$ native US speakers to \textit{'Please choose the better sounding version.'}. The in-domain results in Table~\ref{tab:preference} show a statistically significant preference for our proposed model  ($p<0.05$ in a binomial test). The preference is found across all expressive domains and informal listening shows that more expressive utterances receive a stronger preference and are characterized by a 'cleaner' sound (i.e. less distortion and buzzing).

\vspace{-0.2cm}
\begin{table}[th]
  \caption{Preference test between Parallel WaveNet with latent vector conditioning and baseline without extra conditioning.}
\vspace{-0.2cm}
  \label{tab:preference}
  \centering
  \begin{tabular}{ r@{}l  r r r r}
    \toprule
    & &
     \multicolumn{2}{c}{\textbf{In-domain}}& 
     \multicolumn{2}{c}{\textbf{Out-of-domain}} \\
    \midrule
    \multicolumn{2}{c}{\textbf{Option}} & 
     \multicolumn{1}{c}{\textbf{Votes}}& 
     \multicolumn{1}{c}{\textbf{\%}} &
     \multicolumn{1}{c}{\textbf{Votes}}& 
     \multicolumn{1}{c}{\textbf{\%}} \\
    \midrule
    &prefer baseline 	& $1301$	&  $9.88\%$ 	& $1345$	&  $26.41\%$     	\\
    &no preference 	& $9445$	&  $71.71\%$ 	& $2152$	&  $42.23\%$       	\\
    &prefer proposed  & $2424$	&  $18.41\%$ & $1598$	&  $31.36\%$      	\\
    \bottomrule
  \end{tabular}
\end{table}
\vspace{-0.5cm}

\subsection{Out-of-domain evaluation}
\vspace{-0.2cm}
We also evaluate our model on another set of data from the same speaker but from an expressive domain unseen during training. The mel-spectrograms are predicted using a separate acoustic model based on the same architecture described in \ref{sec: prosotron}. We evaluate $100$ utterances by asking $50$ native US speakers to \textit{'Please choose the better sounding version.'}. The out-of-domain results in Table~\ref{tab:preference} again show a statistically significant preference for our proposed model  ($p<0.05$ in a binomial test). 

\vspace{-0.2cm}
\subsection{Different speaker evaluation} \label{sec: scaling}
\vspace{-0.2cm}
To understand if the proposed approach scales to other voices, we train the same acoustic and Parallel WaveNet models on $10$~hours of expressive data of a male US voice. We apply our proposed technique and run a MUSHRA evaluation using 200 utterances comparing Parallel WaveNet with and without the additional conditioning. We ask 6 native US speakers to rate the systems in terms of their naturalness. The recordings are rated $\textbf{97.48}$, while the model without extra conditioning received $\textbf{47.70}$ and the model with extra conditioning $\textbf{58.97}$. The p-values of pairwise two sided t-tests between all systems are 0. The additional conditioning on latent vectors improves the MUSHRA score by 11 points, showing that the approach can be scaled to other expressive voices. The gap between both synthesis systems and recordings is large since it is a very expressive voice with fairly limited amount of data resulting in less accurate predictions from the acoustic model. The results indicate that the additional conditioning can bring a larger uplift on overall lower quality voices.

\vspace{-0.3cm}
\section{Discussion \& interpretation} \label{sec: analysis}
\vspace{-0.1cm}
We have shown that conditioning the Parallel WaveNet models on the latent vector helps with synthesis quality in different scenarios. Since the baseline Parallel WaveNet models are already conditioned on the spectrograms, we need to understand why conditioning them on a latent vector derived from the same spectrograms brings an improvement. Our hypothesis is that the latent vectors represent a compressed version of the spectrograms in a bottlenecked space in which there is less difference between natural and predicted spectrograms. The vocoder is conditioned on natural spectrograms during training and on predicted ones during inference. The extra conditioning provides some additional information about corresponding  spectrograms, reducing the mismatch between training and inference.

\vspace{-0.2cm}
\subsection{Natural speech} \label{sec: scaling}
\vspace{-0.2cm}
To test our hypothesis, we perform an evaluation on spectrograms from natural speech rather than predicted ones. We focus on the long-form domain only and evaluate $100$ utterances from a book unseen during training, asking $30$ native US listeners to rate the systems on a scale from $1$ to $100$. We perform two tests, one asking the listeners to \textit{'Please rate the systems in terms of their naturalness.'} and one asking them to \textit{'Please listen to the reference sample on each testing screen and then rate how similar the systems sound to the reference audio.'} providing the recordings as reference audio. The results are shown in the top half of columns 1) and 2) in Table~\ref{tab:mushra}. For both questions, a pairwise two sided t-test shows no statistical difference between baseline and proposed system (p-value $>0.05$). The p-values between recordings and both systems are $0.0$. The evaluation on corresponding predicted spectrograms shows a statistically significant preference for the proposed system (see top half of column 3) in Table~\ref{tab:mushra} and section \ref{sec: scaling} for more details).

\begin{table}[th]\
  \caption {MUSHRA test results for predicted and natural speech spectrograms comparing vocoders without additional conditioning (baseline) to ones conditioned on different latent vectors.}
\noindent\makebox[\columnwidth]{
  \vspace{-0.2cm}
    \small
    \begin{tabular}{l c c |c}
    \toprule
    \textbf{System} & \multicolumn{3}{c}{\textbf{MUSHRA test}} \\
    \midrule
      & 1) Natural & 2) Natural & 3) Predicted \\
     &  &  \hspace{10pt} Similarity &\\
    \midrule
    Recordings  & $74.40$ & N/A & N/A \\
    Baseline    & $71.57$ & $72.09$ & $64.09$ \\
    Proposed    & $71.80$ & $72.04$  & $\textbf{65.42}$ \\
    \midrule
    VAE long-form   & $72.15$ & $71.68$ & $\textbf{65.41}$ \\
    VAE neutral   & $72.16$ & $71.81$ & $\textbf{65.66}$ \\
    VAE random   & $71.53$ & $72.30$ & $\textbf{65.06}$ \\
    \bottomrule
    \end{tabular}
    }
    \label{tab:mushra}
\vspace{-0.2cm}
\end{table}

These results indicate that the additional conditioning helps the model to produce better quality only when vocoding predicted spectrograms. To get a better understanding of the relation between natural and predicted speech in the latent space, we calculate the KLD between clusters as a measure of distances. We assume Gaussian distributions for the different clusters of expressive data (neutral, news, long-form book reading, emotions) and calculate the two-sided averaged KLD between all domain clusters. Table~\ref{tab:kld} shows that for long-form the natural and predicted clusters are closer to one another than the natural long-form cluster is to any other natural cluster and the predicted long-form cluster is to any other predicted cluster. The results for the other clusters are not shown but follow the same trend. This further supports our hypothesis that the latent vectors of predicted spectrograms carry additional information about corresponding natural spectrograms.

\vspace{-0.3cm}
\begin{table}[th]\
  \caption {KLD between different distributions in the latent space.}
\noindent\makebox[\columnwidth]{
\vspace{-0.3cm}
    \small
    \begin{tabular}{l c c}
    \toprule
     & natural & predicted  \\
     & long-form & long-form  \\
    \midrule
    natural long-form  & $\textbf{0}$ & $\textbf{1}$  \\
    natural emotions   & $8$ & $8$  \\
    natural news   & $15$ & $15$  \\
    natural neutral  & $14$ & $14$  \\
    \midrule
    predicted long-form   & $\textbf{1}$ & $\textbf{0}$  \\
    predicted emotions  & $8$ & $7$  \\
    predicted news  & $61$ & $57$  \\
    predicted neutral   & $13$ & $11$  \\
    \bottomrule
    \end{tabular}
    }
    \label{tab:kld}
    \vspace{-0.5cm}
\end{table}

\subsection{Inference conditioning} \label{sec: scaling}
\vspace{-0.2cm}
To better understand the importance of the latent vector conditioning during inference, we provide latent vectors other than the one extracted from the predicted spectrogram. We work with the same test set as in section~\ref{sec: scaling} but with predicted spectrograms instead of ones extracted from natural speech. The predicted spectrograms are created using an acoustic model different from the one repurposed as the latent vector extractor but also based on the same architecture including a VAE reference encoder, trained on neutral and long-form data. We are comparing the following latent vectors: 

\begin{itemize}
\item \textbf{Baseline}: without additional conditioning.
\vspace{-0.1cm}
\item \textbf{Proposed}: latent vector extracted from the spectrogram.
\vspace{-0.1cm}
\item \textbf{VAE long-form}: latent vector mean of the distribution of training utterances in the long-form domain.
\vspace{-0.1cm}
\item \textbf{VAE neutral}: latent vector within the distribution of the neutral domain of the training data.
\vspace{-0.1cm}
\item \textbf{VAE random}: a random latent vector close to but outside of the training distribution.
\item \textbf{VAE outside}: a latent vector several magnitudes outside of the training distribution.
\vspace{-0.1cm}
\end{itemize}

\begin{figure}[t]
  \centering
  \includegraphics[width=\linewidth]{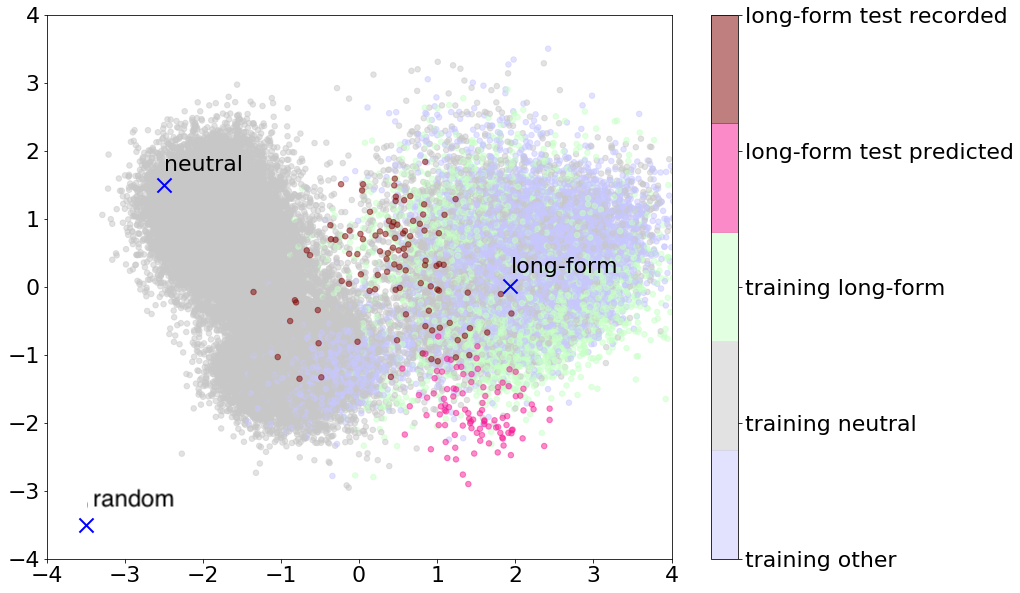}
\vspace{-0.7cm}
  \caption{PCA projection of the VAE latent space. Xs mark the positions used for conditioning.}
\vspace{-0.5cm}
  \label{fig:pca}
\end{figure}

Figure~\ref{fig:pca} shows a two-dimensional PCA projection of the VAE latent space created by the latent space extractor. Blue Xs indicate the latent vectors used for conditioning. Grey points represent the latent vectors of neutral training data, light green training points in the long-form domain and light blue training points from other expressive domains. The brown and pink points indicate the extracted latent vectors for the utterances used in the evaluations; the recordings and their predicted counterparts respectively.

We evaluate the synthesis quality using a MUSHRA tests, asking $30$ native US listeners to\textit{'Please rate the systems in terms of their naturalness.'} on a scale from $1$ to $100$. Using the VAE outside vector, the synthesized audio only contained noise and therefore we did not include this in the evaluation. The results of the MUSHRA tests can be seen in column 3) of Table~\ref{tab:mushra}. The vocoders conditioned on latent vectors are rated  higher than the baseline, by between $1$ and $1.5$ points. The p-values of a pairwise two sided t-test between the baseline and all VAE systems are $<0.05$, showing statistical significance of the results. The p-values between all VAE systems are $>0.05$, indicating that there is no difference in naturalness when conditioning on different latent vectors.

These results show that providing the additional conditioning during training helps to create a model that produces better quality when vocoding predicted spectrograms. Once the model is trained, the actual vector used for conditioning during inference does not seem to affect the quality. However, unreasonable points far outside of the training distribution completely destabilize the model and result in pure noise inference. This finding allows us to train an improved vocoder by providing additional conditioning during training while not having to introduce additional latency during inference for the computation of latent vectors. We are able to provide the mean of the training distribution and still get improved synthesis quality compared to the baseline.

\vspace{-0.3cm}
\section{Conclusions and future work} \label{sec: conclusions}
\vspace{-0.1cm}
We have investigated the use of a global conditioning vector to boost the signal quality of a Parallel WaveNet neural vocoder. This is achieved by conditioning the neural vocoder with the latent vector from a pre-trained VAE component from a Tacotron 2-style sequence-to-sequence model, which summarises the spectrogram sequence at an utterance-level. We find that by conditioning the Parallel WaveNet model with the latent vector we are able to significantly improve the signal quality when synthesizing predicted spectrograms, a conclusion which was confirmed on a second very expressive speaker. Our investigations indicate that the latent vector informs the model during training to the region of speech, helping to bridge the gap between predicted spectrograms and those extracted from natural speech. The selection of the latent vector to use at synthesis-time appears not to be important, providing it comes from a point which is close to that which was observed in the training data. This indicates that we are able to make use of the improvements of this model without needing to introduce latency during inference for the computation of latent vector. We are able to provide the fixed mean of the training distribution to condition the model.

In this paper we investigated the use of a pre-trained VAE component for conditioning the Parallel WaveNet model; however, there are a number of questions which remain to be investigated in future work. Firstly, does the component need to be pre-trained? Is this conditioning acting as an expert which informs the Parallel WaveNet model of the behaviours of the \textit{speech space}, or is it providing the model with an utterance-level approximation and prior knowledge is not necessarily required, i.e. this utterance-level representation can be jointly trained with the network? Secondly, is this conditioning best handled in the form of a variational representation or can other global summarizations of the spectrogram sequence be equally effective? Finally, in this investigation we focused on a speaker dependent neural vocoder. Previously we found that we are able to train `universal' WaveRNN-style neural vocoders to generate high quality speech which is invariant to a number of unseen scenarios \cite{Lorenzo-Trueba2018}. Future work is required to understand whether we are able to develop a universal Parallel WaveNet neural vocoder and whether the global conditioning vector also brings a performance gain in such a scenario.

\vfill\pagebreak

\bibliographystyle{IEEEbib}
\bibliography{mybib}

\end{document}